# Behaviors of QCA Inverter due to Cell Displacement and Temperature Variation

**Angshuman Khan[1*], Surajit Sur[1], Chiradeep Mukherjee[1], Aninda Sankar Sukla[1], Ratna Chakrabarty[2]**

[1]Department of Electronics & Communication Engineering, University of Engineering & Management, Jaipur, Rajasthan-303807, India
[2]Department of Electronics & Communication Engineering, Institute of Engineering & Management, Kolkata, West Bengal-700091, India
* Corresponding author(s): Angshuman Khan; angshumankhan2910@gmail.com

**Abstract**

*Quantum dot Cellular Automata (QCA) is the emerging area in the field of nanotechnology. Inverter is a fundamental logic primitive in QCA. Molecular, semiconductor, magnetic, and metallic QCA are main methodology in the fabrication of quantum cell. While all types of QCA work on room temperature, metallic one is not suitable in normal temperature. So temperature plays a significant role in QCA circuit. In this paper, the effect of temperature in two-cell conventional inverter and recently proposed three-cell high polarized inverter has been discussed. The polarization and Kink energy of QCA circuit is influenced due to the change of distance between two cells. This paper clearly mentioned the variation of polarization and kink energy of QCA inverter due to cell displacement. Finally this paper makes a comparison between the conventional two-cell inverter and recently proposed three-cell inverter. The simulation tool QCADesigner has been used to study the effects of QCA.*

## I. INTRODUCTION

Quantum dot Cellular Automata (QCA) is the most promising nanotechnology in this VLSI era. It could make digital circuits with very low feature size in nanometer range. It has extremely low power consumption [16]. In QCA technology, there are some basic important logic primitives for any digital design purpose. Inverter is one of them. For this reason, inverter is the experimental object in this article. Among the many conventional QCA inverters, the two cells inverter is the smallest one, but it has low polarization. The recently proposed [10] high polarized QCA inverter consists of three cells. These two inverters are the investigational entity in this paper as they are very popular building blocks of QCA technology. There are four main approaches to fabrication of QCA cell, named molecular, semiconductor, magnetic, and metallic [2], [8]. The molecular, semiconductor and magnetic QCA work on normal temperature, but metallic QCA is not appropriate in room temperature. Temperature being an important parameter, it plays a crucial role in QCA circuit designing. In this paper, the effect of temperature in two-cell conventional inverter and three-cell high polarized inverter [10] has been discussed. The polarization of QCA circuit is prejudiced due to the change of distance between two cells. This paper clearly mentioned the effect of displacement of output cell in these two types of inverters. Not only the polarization but the cell displacement can alter the gesture of kink energy and hence the stability. The entire argument leads to the fact that cell displacement links kink energy with circuit's stability. Finally this paper concludes with a comparison between the two small inverters, that is the two-cell state-of-the-art inverter and recently proposed high polarized three-cell inverter typified on temperature and cell displacement on polarization, kink energy and stability. The simulation tool QCADesigner has been used austerely to study the circuits in this paper.

## II. QCA Background

### A. Physics of QCA

The heart of QCA technology is a QCA cell. It consists of four quantum dots. This four quantum dots of a QCA cell, located at corner position with two extra mobile electrons. A QCA cell has two polarization arrangements. These two arrangements of polarizations are denoted as cell polarization P = +1 (binary "1") and P = -1(binary "0"), as shown in Fig. 1. In each polarization state, electrons tend to capture the opposite locations for their mutual electrostatic repulsion [2]-[8].

### B. Elementry Logic primitives

The basic three building blocks of QCA logic devices include QCA majority gate, QCA wire, and QCA inverter [2]-[9].

*QCA Wire:* QCA wire is an array of QCA cells on a line. The wire is driven at the input cell by a fixed (held) polarized cell [2]-[8]. Fig. 2 shows a wire with input and output is "1".

*Majority Gate:* The logic function of a majority gate is M (A, B, C) = AB + BC + CA, where A, B, and C are arbitrary inputs. Generally, majority gate is a three inputs logic function as shown in Fig.3. By keeping the polarization of any input of the majority gate as logic "1" or logic "0", an OR gate or AND gate will be obtained like: M (A, B, 0) = AB, and M (A, B, 1) = A+B [2]-[8].

*QCA Inverter:* There are different types of QCA inverters have discovered yet, but the two-cell state-of–the-art inverter is very popular due to its small size, though it has a low polarization. The diagram of a conventional two-cell QCA inverter circuit and recently proposed high polarized QCA inverter circuit according to [10] is shown in Fig. 4 and Fig. 5, respectively.

*Clocking Scheme:* Clock in QCA system used to define the direction of state transition. QCA circuit areas are divided into four clocking zones, named Switch, Hold ,Release and Relax [2],[8] as shown in Fig. 6. During the Switch phase, the inter-dot barriers are slowly raised and according to the state of their drivers (that is, their input cells), the QCA cells become polarized. The output of this state can be used as the inputs to the next state. During the Hold phase, the inter-dot barriers are kept high and the QCA cells retain their present states. In the Release phase, the barriers are lowered and the cells are allowed to relax and they enter in an un-polarized state. Finally, in the Relax phase, the barriers are kept low and the cells remain un-polarized. As shown in Fig. 6, always there is a 90 degree phase shift from one clock zone to the next. In each clock zone, the clock signal has four states, named: high-to-low, low, low-to-high, and high [10]-[15].

## III. Kink Energy

The electrostatic interaction occurs between two QCA cells [9]-[13], [15] and it is given by equation (1)

$$E = \frac{1}{4\pi\epsilon_0\epsilon_r} \cdot \frac{Q_1 Q_2}{r} = k \cdot \frac{Q_1 Q_2}{r} \quad \text{...........................(1)}$$

Where, the value of "k" is $9 \times 10^{-9}$ and $Q_1$ and $Q_2$ are charges of electron. So, the figure of "E" follows the equation (2). This electrostatic interaction determines the Kink energy between two cells. The Kink energy is defined as equation (3). A point to be remember, the kink energy between two QCA cells only depends on the dimensions of the QCA cells and the spacing between the cells but it does not depend upon the temperature [6], [9].

$$E = \frac{23.04 \times 10^{-29}}{r} \quad \text{.......................................(2)}$$

$$E_{kink} = E_{opp.polarization} - E_{same\ polarization} \quad \text{.....(3)}$$

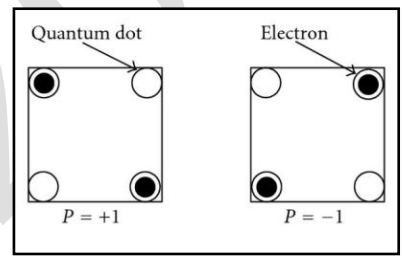

Fig. 1: QCA cells

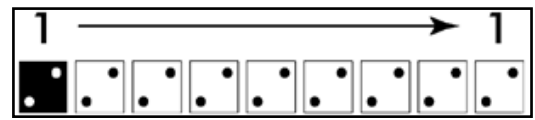

Fig. 2: QCA wire

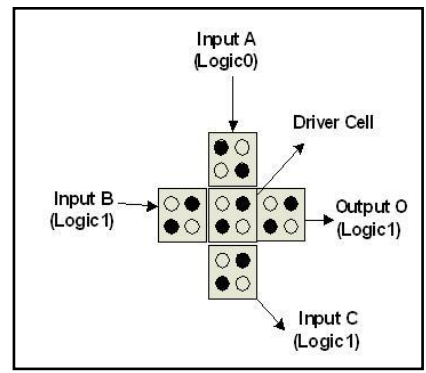

Fig. 3: Majority logic

2 nm for all implemented inverters. There is no remark on their kink energy, hence stability.

## VI. PRESENT WORK

### A. Temperature Vs. Polarization

Temperature has a significant effect in the polarization of QCA cell. On the contrary, polarization of inverter may change due to the change of temperature. As we know, the metal island implanted QCA doesn't work on room temperature, while semiconductor, molecular, and magnetic type QCA may work at ordinary temperature. This paper investigates the effect of temperature in conventional two-cell inverter and recently proposed three-cell inverter. Table I shows the change of polarization in different temperature for both type of inverters simulated in QCADesigner tool. The plot for change of polarization with respect to temperature is shown in Fig. 7.

### B. Displacement vs. Polarization

Not only is the temperature, distance an important issue in polarization. Distance between two cells determines the polarization of a cell. In this article, the polarization of inverters is examined in QCADesigner tool in different displacement of the output cell from its previous cell. Table II clearly mention the change of polarization in different distance of output cell for both type of inverter, state-of-art two-cell inverter and only just proposed three-cell inverter. Fig. 8 shows the plot for displacement of output cell with respect to its previous cell versus polarization.

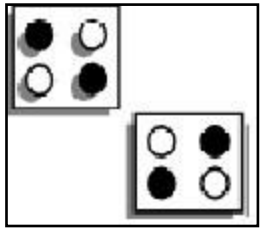

Fig. 4: Conventional inverter

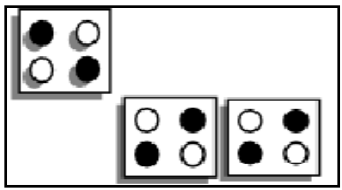

Fig. 5: Recently proposed high polarized inverter according to [10]

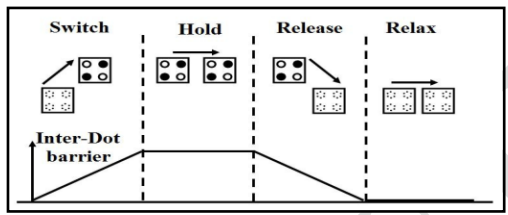

Fig. 6: QCA clocking

## IV. QCADESIGNER

All the circuits in this paper have been checked in QCADesigner tool, with version 2.0.3[1]. The default parameters for coherence vector mode are:
Temperature: 1K
Relaxation Time: 1.000000e-015 s
Time Step: 1.000000e-016 s
Total simulation Time: 7.000000e-011 s
Clock High: 9.800000e-022 J
Clock Low: 3.800000e-023 J
Clock Shift: 0.000000e+000 J
Clock Amplitude Factor: 2.000000
Radius of Effect: 80.000000 nm
Layer Separation: 11.500000 nm

## V. PREVIOUS WORK

All introduced inverter models including three-cell newly designed inverter are implemented in 1K. The spacing between the output cell and its previous cell is

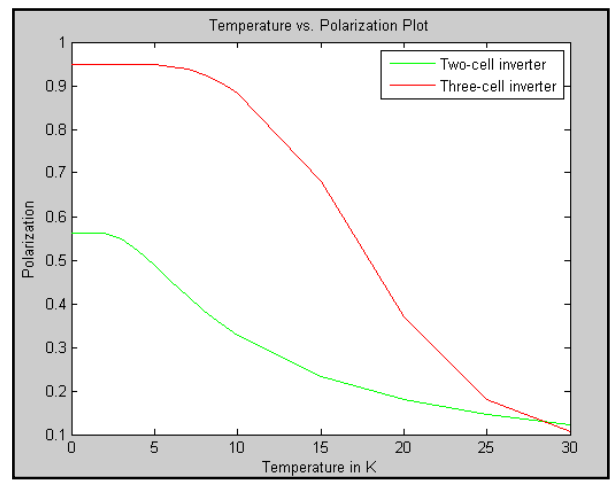

Fig. 7: Temperature vs. Polarization plot

TABLE I        TEMPERATURE VS. POLARIZATION

| Temperature | Polarization | |
|---|---|---|
| (In K) | Two-cell (conventional) | Three-cell (newly proposed) |
| 0 | 0.562 | 0.950 |
| 1 | 0.562 | 0.950 |
| 2 | 0.561 | 0.950 |
| 3 | 0.549 | 0.950 |
| 4 | 0.523 | 0.950 |
| 5 | 0.489 | 0.948 |
| 6 | 0.452 | 0.944 |
| 7 | 0.416 | 0.937 |
| 8 | 0.383 | 0.924 |
| 9 | 0.354 | 0.907 |
| 10 | 0.327 | 0.884 |
| 15 | 0.234 | 0.679 |
| 20 | 0.181 | 0.370 |
| 25 | 0.146 | 0.181 |
| 30 | 0.123 | 0.107 |

TABLE II        DISPLACEMENT VS. POLARIZATION

| Distance between output cell and its previous cell | Polarization | |
|---|---|---|
| (In nm) | Two-cell (conventional) | Three-cell (newly proposed) |
| 0.50 | 0.610 | 0.973 |
| 1.00 | 0.566 | 0.968 |
| 1.50 | 0.560 | 0.958 |
| 2.00 | 0.562 | 0.950 |
| 2.50 | 0.535 | 0.935 |
| 3.00 | 0.510 | 0.908 |

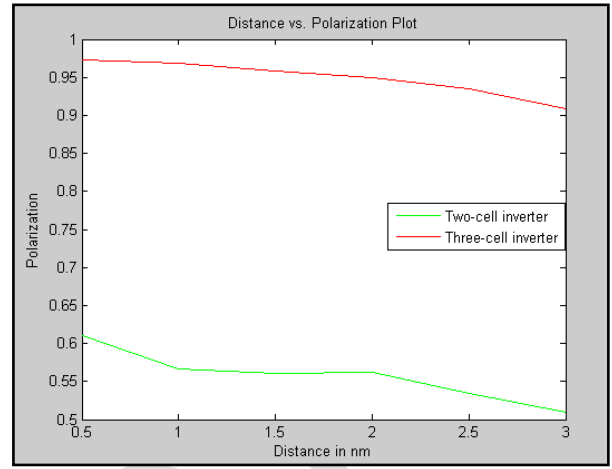

Fig. 8: Distance vs. Polarization plot

### C. Displacement Vs. Kink Energy

We know that, temperature doesn't effect on the kink energy of QCA cell. The kink energy between two cells depends upon the dimension of the cell and the distance between the cells. In this text, the change of kink energy has been noticed due to the change of output cell of QCA inverter. Table III shows the details change of kink energy due to change of distance in case of inverter and Fig. 9 shows the corresponding plots.

TABLE III        DISPLACEMENT VS. KINK ENERGY

| Distance between output cell and its previous cell | Kink energy (in Jules) | |
|---|---|---|
| (In nm) | Two-cell (conventional) | Three-cell (newly proposed) |
| 0.50 | $3.394 \times 10^{-20}$ | $44.162 \times 10^{-20}$ |
| 1.00 | $3.386 \times 10^{-20}$ | $21.249 \times 10^{-20}$ |
| 1.50 | $3.3486 \times 10^{-20}$ | $13.518 \times 10^{-20}$ |
| 2.00 | $3.364 \times 10^{-20}$ | $9.714 \times 10^{-20}$ |
| 2.50 | $3.354 \times 10^{-20}$ | $7.445 \times 10^{-20}$ |
| 3.00 | $3.340 \times 10^{-20}$ | $5.945 \times 10^{-20}$ |

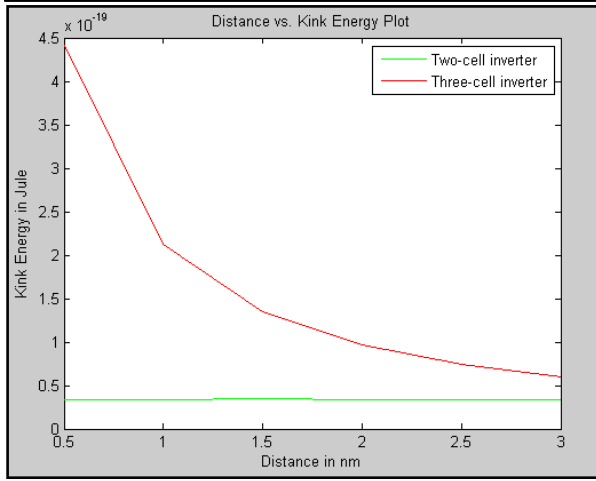

Fig. 9: Distance vs. Kink energy plot

## VII. COMPARISON

For both the inverters, the polarization increases with the increase of temperature. There is a drastic change of polarization after 10K-15K in case of both inverters. There are no changes of polarizations in the range of temperature 0K-4K in three-cell newly proposed inverter. In higher temperature (after 30K), polarization is almost same for both inverters.

Not only the temperature, the distances between cells have also important effect in polarization. For both inverters, with the decrease of distance between the output cell and its previous cell, the polarization increases and vice versa. But the recently proposed three-cell inverter has higher response at an equal spacing with respect to conventional two-cell inverter.

The temperature doesn't effect in the kink energy of QCA circuits. Kink energy only depends upon the cell dimensions and distance between cells. Higher kink energy indicates the instability of QCA circuits. The kink energy doesn't affected significantly with the displacement of output cell in case of two-cell inverter, hence indicates its stability. But at a lower spacing between output cell and its previous cell, there is a drastic change of kink energy in case of three-cell newly proposed inverter. Very high kink energy at lower distance between cells means the instability of circuit. With the increase of cell distance, the kink energy increases for this three-cell inverter, hence stability decreases.

## VIII. CONCLUSION

Quantum dot Cellular Automata attracts researcher to focus on due to its ultra low power consumption, and tremendous speed. One of the fundamental elements of QCA technology is inverter which has been focused in this article. There are many QCA inverters implemented yet. Among them two cell state-of-art inverter is famous due to its small size, but it is not well polarized. There is another inverter proposed recently with high polarization. It has three cells, instead of two. These two inverters are the experimental circuits in this paper. In lower temperature, there is no change of polarization in for three-cell inverter. Hence it is more suitable in lower temperature with respect to two-cell inverter whose polarization changes with the change of temperature in low temperature ranges. At higher temperature, the polarizations are almost same for both inverters; so, two-cell inverter is more preferable due to size issue. The cell displacement certainly affects in the response of two-cell inverter, but there is a drastic change in output in three-cell inverter at lower cell distance. So, two-cell inverter is stable in any displacement of cells. But at very low cell distance of three-cell inverter, it has very high kink energy, it is not stable here. But at higher distance of cells, the kink energy of three-cell inverter is low, so it is very stable for that case.

Finally the conclusion is that both inverters have its own advantages as well as disadvantages. But the recently proposed three-cell inverter is more preferable because of its high polarization, moderate size, stability at lower temperature, and correctness for higher distances of cells.